\documentclass{epl}

\title{Friction Laws for Elastic Nano-Scale Contacts}

\author{L. Wenning \and M. H. M\"user}

\institute{Inst. f. Physik, Johannes Gutenberg-Universita\"at,
           55099 Mainz, Germany}

\begin{document}

\maketitle

\begin{abstract}
The effect of surface curvature on the law relating frictional
forces $F$ with normal load $L$ is investigated by molecular dynamics
simulations as a function of surface symmetry, adhesion, and
contamination. Curved, non-adhering, dry, commensurate surfaces show a linear
dependency, $F \propto L$, similar to dry flat commensurate or amorphous
surfaces and macroscopic surfaces.
In contrast, curved, non-adhering, dry, amorphous surfaces
show $F \propto L^{2/3}$ similar to friction force microscopes.
In our model, adhesive effects are most adequately described
by the Hertz plus offset model, as the
simulations are confined to small contact
radii. Curved lubricated or contaminated
surfaces show again different behavior; details depend on
how much of the contaminant gets squeezed out of the contact.
Also, it is seen that the friction force in the lubricated case
is mainly due to atoms at the entrance of the tip.
\end{abstract}

\section{Introduction}
Amontons law, which connects the frictional force $F$ between two solids
in relative motion linearly with the normal load $L$, was suggested more
than 300 years ago.  This law has proven applicable since for many sliding 
interfaces~\cite{dowson79,persson98}.
Nevertheless, it is still discussed controversially whether or not Amontons'
law is valid on the micrometer or the nanometer scale as well, e.g., whether
it holds for individual asperity contacts.
If a contact deforms plastically, there is a simple popular, yet,
phenomenological argument why this should be the case~\cite{baumberger96}:
The local normal pressure $p_\perp$ in the contact is everywhere
close to the yield pressure $p_{\rm y}$ and the shear stress of
the junction is limited through the yield stress $\sigma_{\rm c}$.
Omitting adhesive effects, this results in a static friction coefficient of
$\mu_{\rm s} = \sigma_{\rm s} / p_{\rm y}$. $\mu_{\rm s}$ is commonly defined
as the ratio of the force $F_{\rm s}$ needed to initiate sliding and $L$.
This argument is of rather limited predictive power for various reasons,
e.g., $p_{\rm y}$ depends strongly on the size of the asperities in 
contact~\cite{persson98}. Furthermore, it can not be applied to elastic, 
wearless friction, which is the subject of many friction force microscope
(FFM)~\cite{mate87,schwarz97,schwarz97a} and surface force apparatus 
(SFA)~\cite{israelachvili92,berman98} experiments. In the following we will
focus on wearless, elastic friction.

Even in the elastic regime, SFA and FFM experiments are often consistent
with the interpretation of a  yield stress $\sigma_{\rm c}$
that is (relatively) independent of the normal pressure $p_\perp$, 
e.g., very careful SFA experiments
observe that $F$ is mainly proportional to the real area of contact 
$A_{\rm c}$~\cite{mcguiggan99}. Similarly, strong deviations from
Amontons' laws are observed in FFM. If the FFM tip has a reasonably
well-defined shape and adhesive forces are included into the load $L$
in terms of a so-called Hertz-plus-offset model, $F \propto L^{2/3}$
is observed~\cite{schwarz97}. This is again consistent with the assumption
of a normal-pressure independent value of $\sigma_{c}$.
However, simulations of friction between flat surfaces of linear
dimensions comparable to FFM contact radii observe Amontons' macroscopic
law~\cite{he99,muser00} or simple generalizations thereof. The relation between
$\sigma_{\rm c}$ and $p_\perp$ is very well described with the
linear relation
\begin{equation}
\sigma_{\rm c} = \alpha \,  (\sigma_{\rm 0} + p_{\perp}),
\label{eq:micro_amon}
\end{equation}
where $\sigma_{\rm 0}$ can be viewed as an adhesive load per
area. $\alpha = \partial \sigma_{\rm c} / \partial p_{\perp}$
can be interpreted as an differential friction coefficient which
is close to $\mu_{\rm s} = \sigma_{\rm c}/p_\perp$ in the case of 
large $p_{\perp}$. 
Support for the validity of Amontons' law on the microscale has also been
provided by SFA experiments in which the adhesive forces between
two mica surfaces had been shielded with an electrolyte solution.
One may conclude that in typical SFA experiments
the term $\sigma_{\rm 0}$ dominates $p_{\perp}$, while in Ref.~\cite{berman98}
where adhesive effects were eliminated, $p_{\perp}$ dominated $\sigma_{\rm 0}$.
Both results are therefore consistent with Eq.~(\ref{eq:micro_amon}) and
thus consistent with the computer simulations.

It remains to be understood why FFM experiments observe a relationship
$F \propto L^{2/3}$. The arguments  used for the SFA experiments which
stresses adhesive effects are not valid for FFM experiments, because the
contact mechanics are qualitatively different. Contacts in SFA experiments
are usually well described 
with the JKR model~\cite{johnson71}, in which a small change in load can
change the adhesive interactions considerably. Hence the effective load is
typically a highly non-linear function of the externally applied load in
an SFA experiment.
Due to the much smaller contact radii, FFM experiments are
described best by the Hertz-plus-offset model~\cite{schwarz96}. 
Here, the adhesive normal forces barely change with the load and simply
give a constant bias to the externally applied normal load.

A recently suggested simple model for the interaction between two
disordered, but atomistically flat surfaces predicts a proportionality
of $\sigma_{c}$ and $p_\perp$. However, the net friction coefficient depends
on the area of contact $A_{\rm c}$ according to
$\mu \propto 1/\sqrt{A_{\rm c}}$ if no contaminating atoms are present on
the surfaces~\cite{muser00}. The predictions of the simple
model are then confirmed by detailed atomistic computer 
simulations~\cite{muser00}. In this letter, we want to show that this
relationship explains the observation of the $F \propto L^{2/3}$ relation
seen in FFM experiments: We may regard adhesive forces as an irrelevant
offset in FFM and neglect them for a moment. The generalization of
the friction law between dry, amorphous, flat surfaces 
$F \propto L/\sqrt{A}$ as suggested in Ref.~\cite{muser00} to curved
surfaces and the
validity of Hertzian contact mechanics for non-adhesive contacts,
in particular $A_{\rm c} \propto L^{2/3}$, automatically lead to
the results seen in FFM experiments. We will test this hypothesis
by means of molecular dynamics simulation and investigate the
friction force of a curved tip on a flat surface as a function
of surface symmetry (commensurate, incommensurate, amorphous),
adhesion (adhesive, non-adhesive) and contamination of air-born
particles, which may alter frictional forces
significantly in both experiment~\cite{martin93} and 
simulation~\cite{he99,muser00,muser00a}.

\section{The model}

\begin{figure}
\twoimages[width=6.5cm]{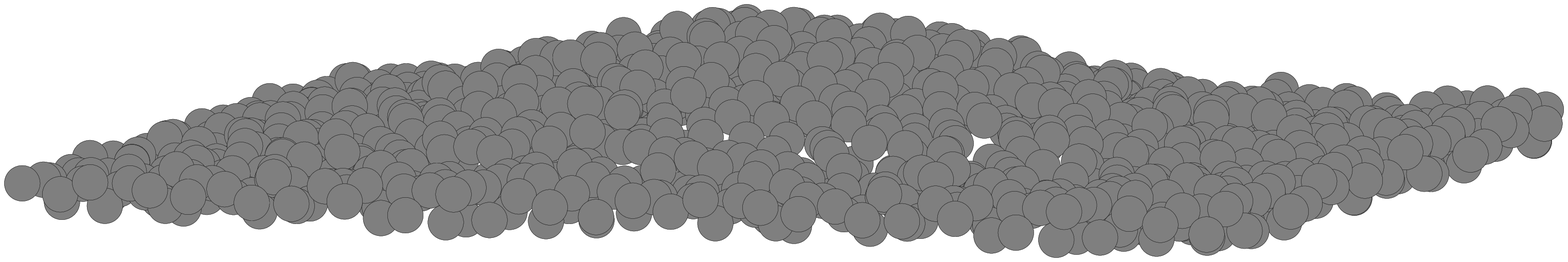}{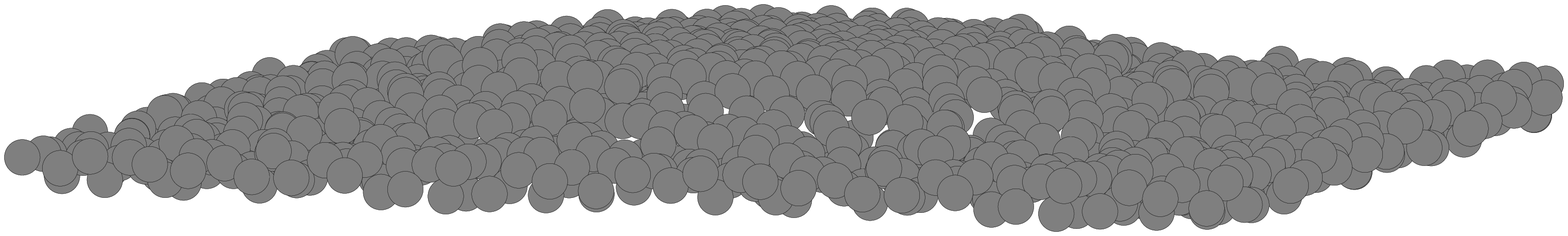}
\caption{Free amorphous tip (left) with $R_{\rm c} = 7$~nm
and the same tip pressed 
against a flat (invisible) surface  with a load of $L \approx 40$~nN (right).
The flattened area can be interpreted as the area of contact.
The tip is shown up-side down in order to improve visualization.}
\label{fig:tip}
\end{figure}

The model used in this study is similar to that used in many 
previous studies, e.g., Refs.~\cite{he99,thompson95}: Atoms in each solid
are coupled elastically to their ideal lattice positions. Interactions
between atoms from opposing walls are Lennard Jones (LJ) interactions.
Additional lubricant atoms (if present) interact with one another and 
with all wall atoms via
LJ potentials, namely $V = 4\epsilon [(\sigma/r)^{12}-(\sigma/r)^{6}]$.
Unlike usually, we will not express the results in reduced LJ units but
rather give them ``reasonable'' dimensions of
$\epsilon = 3\times 10^{-21}$~J and $\sigma = 3.5\,\AA$.
There is one important new feature in the model, which allows for
long-range elasticity orthogonal to the interface.
The basic idea is similar to the one used in a recent computer simulation
study of the squeezing out of lubricants between a curved and a flat
surface~\cite{persson00}.
In this letter, the tip atoms are coupled to their ideal, stress-free positions
in such a way that the ideal Hertzian pressure profile
is generated if the tip is pressed
on a perfectly flat, infinitely hard, and non-adhering surface.
This can be achieved simply by using a normal restoring force 
\begin{equation}
f(\delta z) = {\sqrt{\delta z} / K \sqrt{R_{\rm c}}}
\end{equation}
with $\delta z$ the normal deflection from the ideal lattice position,
$R_{\rm c}$ the radius of curvature of the tip, and $K$ the 
bulk modulus of the tip, which was typically chosen
$B = 25$~GPa. The equilibrium positions are 
given by a function that is sinusoidal in $x$ and $y$ direction
and periodic with the length of the simulation box $L_{\rm box}$.
$R_{\rm c}$ is therefore a function of the height $h$ of the tip and 
$L_{\rm box}$.  A typical amorphous tip is 
is shown in Fig.~\ref{fig:tip} at zero and maximum externally applied
load. The pressure free tip has a height of about $h = 11\,\AA$.
The lateral linear dimension of the simulation box shown is 125~$\AA$.

\section{Results}
\begin{figure}
\onefigure[width=8cm]{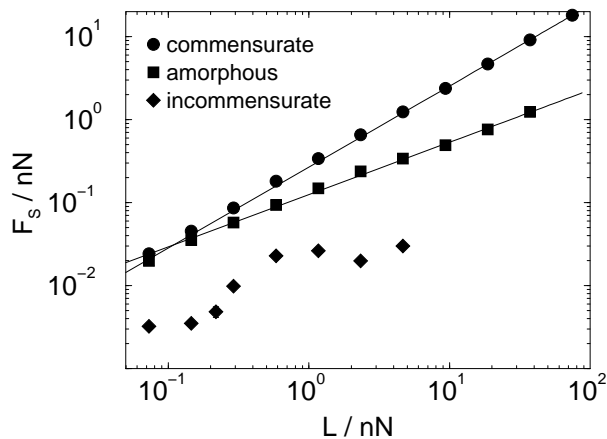}
\caption{Static friction force $F_{\rm s}$ vs. normal load $L$ for
a commensurate tip, an incommensurate tip, and an amorphous tip.
In all three cases, the contact radius was $R_{\rm c} = 70\,\AA$.
The contacts were non-adhesive. Straight lines are fit according
to $F_{\rm s} \propto L^\beta$.}
\label{fig:dry}
\end{figure}

We first consider the case of dry surfaces and suppress the effect
of thermal activation by chosing extremely small temperatures.
Furthermore, adhesive effects between the walls are eliminated by
cutting off the LJ potential in the minimum. Three different tips
are simulated, (i) one tip commensurate with the substrate,
(ii) the same tip as in (i) but rotated by 90$^o$ resulting in
incommensurability, and (iii) a tip with a disordered structure
which we call amorphous. Eliminating thermal activation and suppressing
adhesion allowed us to observe the load-friction behavior down to
very low normal loads with large resolution, i.e.,  we know the
pull-off force to be exactly zero.
The static friction force $F_{\rm s}$
vs. load $L$ curves are shown in Fig.~\ref{fig:dry}.

The results can be interpreted as follows:
(i) For commensurate surfaces, all tip atoms in the contact are basically
``in phase'', that is to say their lateral positions with respect to
the underlying substrate are identical (modulo the lattice constant).
From flat commensurate surfaces, we have learned that
on the atomic scale $F_{\rm lateral}^{\rm (atomic)} \propto L^{\rm (atomic)}$
is rather well satisfied~\cite{muser00}.
The linearity $F \propto L$ for commensurate
tips is therfore not surprising. However, care has to be taken.
This analysis is premature: Depending on the local normal pressure, atoms are
deflected differently in their lateral positions. If we plot individual
contributions to the friction force as a function of the load that individual
atoms carry, no linearity is observed, but rather 
$f_i \propto l_i^\beta$ with $f_i$ the lateral force and $l_i$ the normal force
that one atom of the tip exerts on the lower wall and $\beta \approx 2/3$. 
Yet, the net friction force $\sum_i f_i$ is
proportional to the net load $\sum_i l_i$. The mechanism how this comes
about is very much related to a Greenwood-Williamson type 
argument~\cite{greenwood66}. 
(ii)
The amorphous tip can be fitted best with a power law $F \propto L^{0.63}$,
which is very similar to the relation predicted in the introduction and
seen in FFM experiments. The sublinear behavior can therefore be explained
with an increasingly geometric misfit between the tip and the substrate
with increasing load.
We have to emphasize that the results shown in
Fig.~\ref{fig:dry} are the result of a statistical average over twelve tips.
Individual force-load curves do typically not result in such smooth
curves. The relative large contact radii in real AFM tips may already lead to
some kind of self-averaging as compared to our simulations.
All our tips have been prepared under identical ``experimental'' conditions, 
however, the tips have been cut out from different regions of an amorphous
solid. 
(iii) 
Incommensurate walls show a systematic increase in $F$ vs. $L$ only
at very small contact radii, where basically one atom is in contact.
As the load is increased, the incommensurability becomes more and more
important and the ratio $F/L$ decreases dramatically. The intermittent
behavior very much depends on the relative number of tip atoms which
are in phase with the substrate. This results in non-trivial 
$\mu_{\rm s}(A_{\rm c})$
behaviour, although the large $A_{\rm c}$ limit is of course zero.

If we allow for adhesion between the surfaces
(by choosing a large cut-off radius), no qualitative changes of the
above stated results can be expected. This can be seen by
calculating the so-called Tabor parameter, which tells us what
contact mechanics model is the most appropriate.~\cite{schwarz97,tabor79}
Even if we use the smallest, physically meaningful value for the tip's
bulk modulus and the largest contact radius tractable with the present
computer resources, we would still be in the Hertz-plus offset regime.

\begin{figure}
\twoimages[width=6cm]{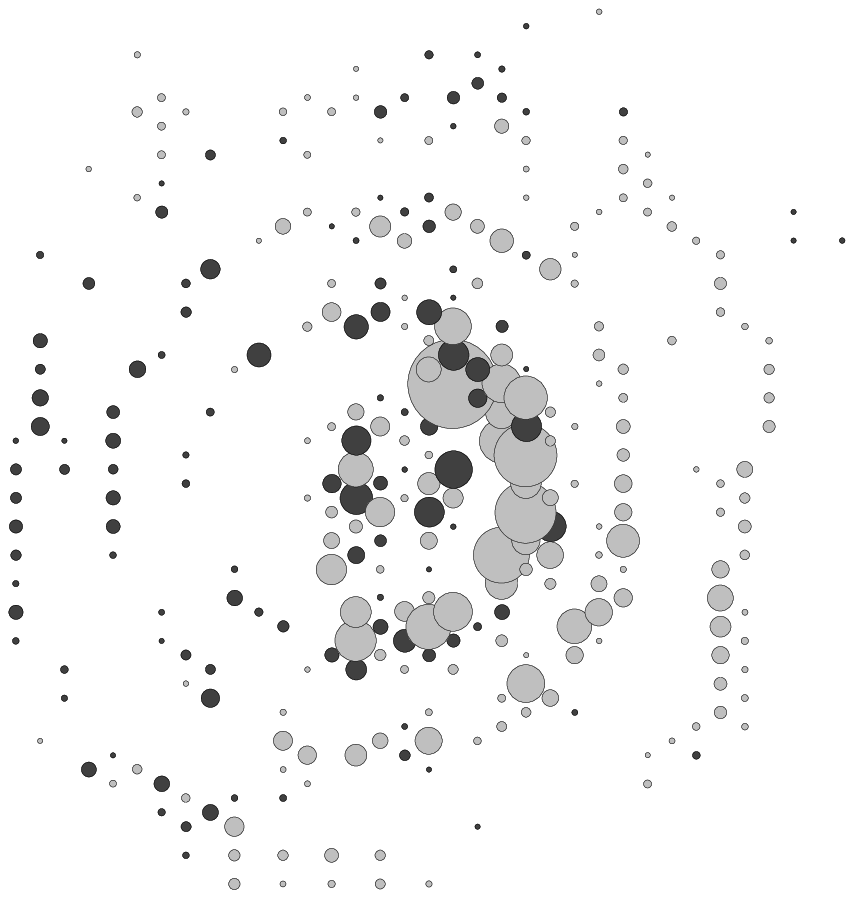}{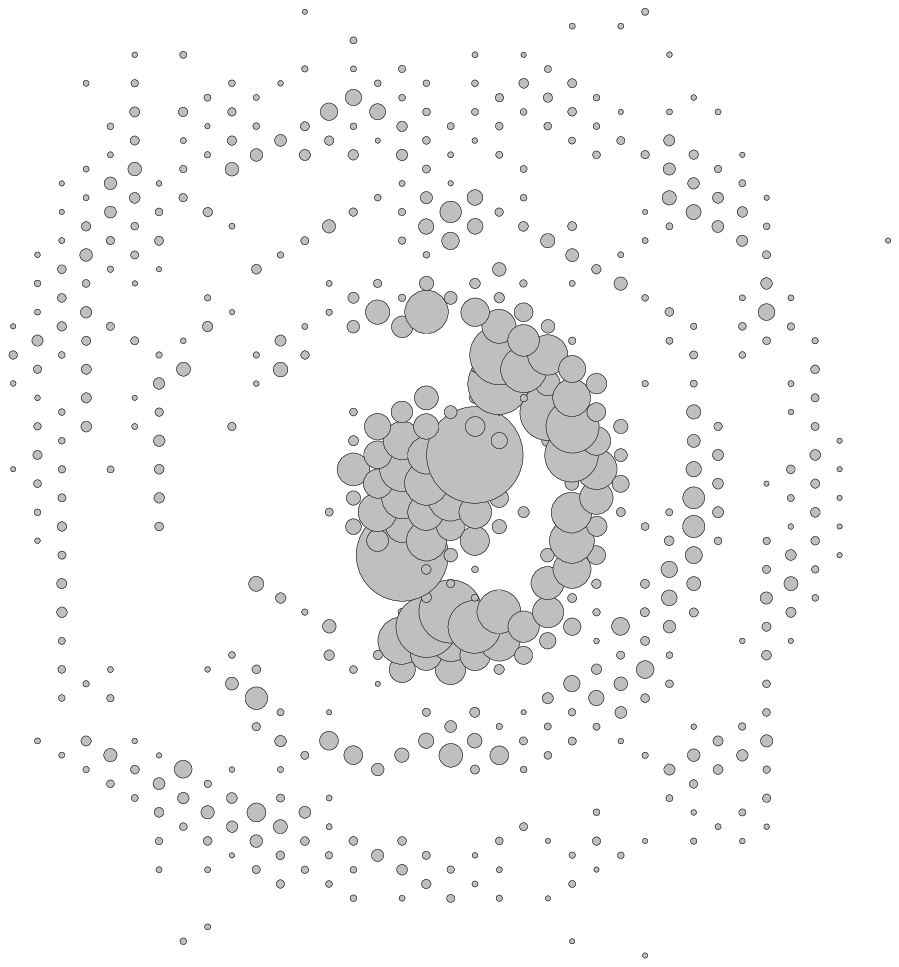}
\caption{Left: Lateral force $f$ on individual tip atoms exerted by the
opposite wall and the lubricant. The size of the particles
is proportional to $| f |$. Light atoms resist the externally applied
force $F_{\rm ext}$, dark atoms support $F_{\rm ext}$.
Right: Normal force $l$ on individual tip atoms exerted by the opposite
wall and the lubricant. The size of the particles is proportional to $l$.}
\label{fig:nor_force}\vspace*{5mm}
\onefigure[width=12cm]{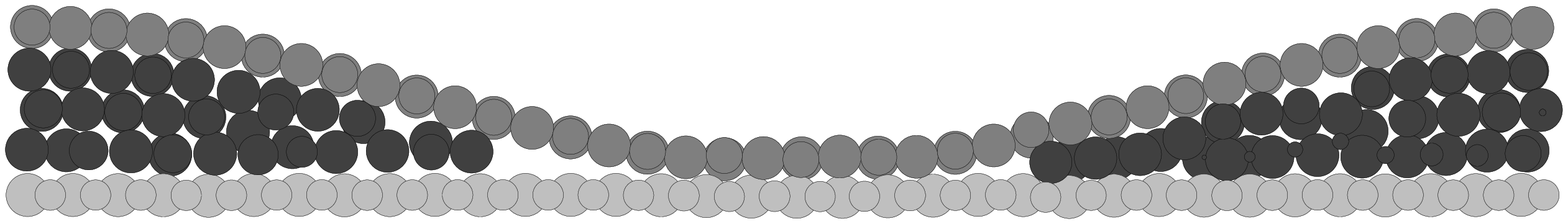}
\caption{Visualization of the lubricated junction. The walls
are incommensurate; the external force acts to the right. In the center,
all of the lubricant (dark atoms) is squeezed out. 
}
\label{fig:junction}
\end{figure}

We now consider the case of physisorbed atoms or molecules on the
surfaces.
These contaminating particles may be air-born or a boundary lubricant.
For a discussion of mechanisms how several layers of lubricant
get squeezed out of the contact see Ref.~\cite{persson00}.
In order to discuss the effect of the physisorbed particles
on the net shear force of the junction,
it is instructive to visualize the distribution of normal
loads and shear forces that individual tip atoms experience.
This is done in Figs.~\ref{fig:nor_force} and \ref{fig:junction}, 
where it is shown which
tip atoms contribute to friction 
and which tip atoms carry the normal load (Fig.~\ref{fig:nor_force}).
In Fig.~\ref{fig:junction}, a cross section of the junction is
visualized. The center of the tip is in direct contact with the
lower wall. There is a large normal pressure in that area, but
the lateral forces are fairly small. More load-carrying
tip atoms can be found  at the center of the ring associated
with a single monolayer lubrication regime. It is noticeable that
only atoms at the entrance contribute to the friction forces, while
atoms at the exit of the tip being located in a geometrically
similar position barely carry any load. The tip atoms at the exit
even experience a force from the film atom
in the direction of the externally applied load. This can change
if adhesive interactions are present as well.
Note that real
surfaces - depending on the chemical composition - may be prone to
cold weld or to generate some kind of debris. Here, however,
we want to confine ourselves to
the case of chemically passivated surfaces and elastic
contacts in order to single out the effect from the lubricant.

\begin{figure}
\onefigure[width=8cm]{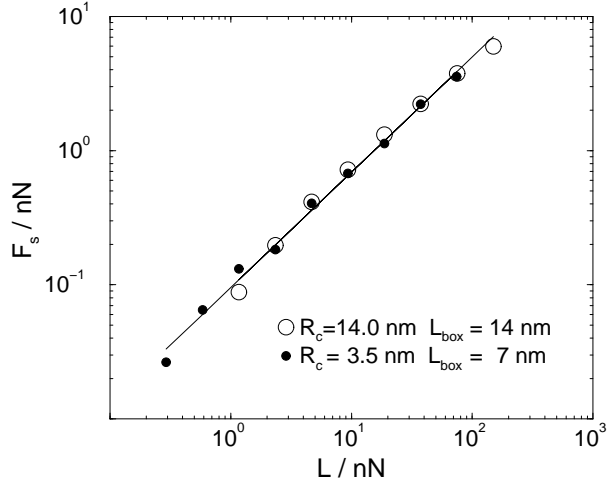}
\caption{Static friction force $F_{\rm s}$ vs. normal load $L$ for
an incommensurate tip on contaminated surface for two different
contact radii $R_{\rm c}$ and length of simulation box $L_{\rm box}$.
The straight line reflects a fit with a power law $F \propto L^{\beta}$
with $\beta = 0.85$.
}
\label{fig:force_lub}
\end{figure}

In the remaining calculations, adhesive effects were again reduced by
making the interaction of tip atoms with substrate atoms and the lubricant
purely repulsive. The only attractive interactions take place between lubricant
and substrate so that the substrate is wetted. We have made sure that
traces in the fluid film on the substrate induced through scraping of the tip
heal sufficiently fast before the tip has been moved for one periodic image.
In Fig.~\ref{fig:force_lub}, force-load curves are shown for a
system consisting of a clean crystalline tip on a contaminated, 
incommensurate surface. 
Sublinear behavior is found, namely $F \propto L^{0.85}$.
This can be interpreted
in such a way that the center of the tip which is in direct contact with
the substrate does not contribute to frictional forces while the
lubricated areas behave similarly as boundary lubricated, flat surfaces
which show $F \propto L$~\cite{he99}. 
The net force results from the superposition of both effects.
Of course, in real experiments the chemical nature of the lubricant will
be relevant as well, i.e., the power law relating $F_{\rm s}$ and $L$
has been observed to depend on the external conditions~\cite{putman95}:
Different laws for a Si$_3$N$_4$ tip on mica were obtained in
argon gas and ambient conditions. 

\section{Conclusions}
This study shows that dry, elastic friction between a curved tip and
a flat substrate can be perfectly understood from the frictional
behavior of dry, flat interfaces: Commensurate systems show linearity
between the friction force $F$ and the load $L$ for both - curved
and flat interfaces. If one of the two
surfaces is amorphous, $F \propto L$ is still valid for a flat interface.
The friction coefficient, however, depends on the area of contact
according to $\mu_{\rm s} \propto 1/\sqrt{A_{\rm c}}$. Assuming
the validity of Hertzian contact mechanics, this leads to a
$F \propto L^{2/3}$ dependency for curved contacts. This has been
confirmed by computer simulations using disordered surfaces and
including the effect of long-range elasticity. The validity of
Hertzian contact mechanics has been imposed by using purely repulsive
potentials. Incommensurate, flat surfaces do not show friction at 
all. For curved surfaces, friction is therefore only seen if 
$A_{\rm c}$ is very small and the ratio $F_{\rm s}/L$ quickly decreases
with increasing contact radius.

Contaminated and lubricated surfaces are more difficult to understand.
For flat surfaces, previous simulations have shown that
Amontons' laws are well satisfied - including the observation that
$\mu_{\rm s}$ does not depend on the area of contact. However, as
explicitly shown for incommensurate tip-substrate systems, very
much depends on the wetting and the squeezing out properties of the
lubricant. At the point of direct contact between tip and substrate,
the systems are not fully equivalent to the dry case, because
the load is not only carried at the center of the tip, but also
further outside, where lubricant atoms decrease the effective
distance between substrate and tip. These ``outskirts'' do not only
carry load but also contribute significantly to the net friction
force. Thus, the friction-load law results from an interplay of
the tribological properties of the direct contact and the (boundary)
lubricated contact. Depending on the details such as wetting and squeezing
properties of the lubricants, different friction-load laws can be expected.
In this study, we have observed a $F \propto L^{0.85}$ dependence
of a boundary-lubricated incommensurate tip-substrate system, which
we do not believe to be universal for this kind of system.

\acknowledgments
We thank K. Binder, B.N.J. Persson, and M.O. Robbins for useful discussions.
M.H.M. is grateful for support through the Israeli-German D.I.P.-Project 
No 352-101.

\end{document}